# An Attack-Driven Incident Response and Defense System (ADIRDS)

**Anthony Cheuk Tung Lai**, VX Research Limited, darkfloyd@vxrl.hk
**Siu Ming Yiu**, University of Hong Kong, smyiu@cs.hku.hk
**Ping Fan Ke**, Singapore Management University, pfke@smu.edu.sg
**Alan Ho**, VX Research Limited, alanh0@vxrl.hk

**Abstract.** One of the major goals of incident response is to help an organization or a system owner to quickly identify and halt the attacks to minimize the damages (and financial loss) to the system being attacked. Typical incident responses rely very much on the log information captured by the system during the attacks and if needed, may need to isolate the victim from the network to avoid further destructive attacks. However, there are real cases that there are insufficient log records/information for the incident response team to identify the attacks and their origins while the attacked system cannot be stopped due to service requirements (zero downtime online systems) such as online gaming sites. Typical incident response procedures and industrial standards do not provide an adequate solution to address this scenario. In this paper, being motivated by a real case, we propose a solution, called "Attack-Driven Incident Response and Defense System (ADIRDS)" to tackle this problem. ADIRDS is an online monitoring system to run with the real system. By modeling the real system as a graph, critical nodes/assets of the system are closely monitored. Instead of relying on the original logging system, evidence will be collected from the attack technique perspectives. To migrate the risks, realistic honeypots with very similar business context as the real system are deployed to trap the attackers. We successfully apply this system to a real case. Based on our experiments, we verify that our new approach of designing the realistic honeypots is effective, 38 unique attacker's IP addresses were captured. We also compare the performance of our realistic honey with both low and high interactive honeypots proposed in the literature, the results found that our proposed honeypot can successfully cheat the attackers to attack our honeypot, which verifies that our honeypot is more effective.

## 1 Introduction

To minimize the financial loss during a cyber-security incident, the incident response team usually follows some industrial standard incident response (IR) frameworks, such as NIST [1] or SANS [2], to investigate the cause of the attack(s). In a typical case, the team mainly relies on system event logs to infer the root cause of the attack(s).

It is not always the case that the team is able to identify the attacks and their origins based on the log records/information provided due to insufficient evidence or imperfect logging systems. In many cases, the victim may need to be disconnected from the infrastructure to avoid further destructive attacks and subject to further analysis. However, there are more and more cases that we cannot shut down or isolate the victim from the infrastructure easily due to service agreement (they are referred as zero downtime online system). Our real case example belongs to this category and is an online gaming site. Note that online gaming industry is one of the top targets for attackers. If company A is attacked, its service is disrupted or there can be system and operational vulnerabilities enabling players to win the game, competitors (of company A) and players can benefit a huge amount from it. The incentive of attacking competitors is high by selling vulnerabilities in the black market (e.g. in dark web).

Other examples include healthcare industry which also has zero downtime systems [13] and critical database [20]. Typical incident response procedures and industrial standards do not provide an adequate solution to address this scenario. In this paper, we propose a solution, called “Attack-Driven Incident Response and Defense System (ADIRDS)” to tackle this problem.

**A real case example**

To facilitate readers to understand the issues, we first briefly present the real case. Our client is a company serving online gaming websites that are accessible to users around the world. These websites process millions of transactions and need to give immediate responses to users. The agreement for service level requirement is extremely high and cannot be taken down (zero downtime online system) without a proper approval from top management and a detailed plan. Because of the uptime requirement, system upgrades and vulnerability patch updates cannot be done frequently. The company only has a single database to store the online gaming data as the production license cost is too high for the company. The followings are the two attacks we finally identified for this case.

**Compromise database transaction records –** Due to insufficient log information, the attack root cause cannot be identified. It is found that the attacker is still able to change the transactions while we carried out our incident response procedure.

**Backdoor deployed in the web server** – A native module has been deployed in the web server, again, no log record is available for to team to detect this backdoor installation.

Thus, the problem is “*how to effectively identify the attacks and their origins to migrate the system from further attacks while keeping the system running with the presence of the attacker*”.

**Our contributions**

In this paper, we propose a solution, called “Attack-Driven Incident Response and Defense System (ADIRDS)” to tackle this problem. ADIRDS is an online monitoring system to run with the real system. By modeling the real system as a graph, critical nodes/assets of the system are closely monitored. Instead of relying on the original logging system, evidence will be collected from the attack technique perspectives (we make use of the list of technical attack activities from an industrial standard of MITRE, called ATT&CK matrix [3]) as our reference to pinpoint the specific evidence to be collected.

To migrate the risks and collect more evidence and footprints from the attacker, we propose a new approach to design realistic honeypots with very similar business context as the real system to be deployed in ADIRDS to trap the attackers. In fact, the community is still hesitated to deploy honeypots in enterprise networks since it may be risky. There are a few other research works that focus on planning, setting up, and deploying honeypots in enterprise networks, such as on-demand virtual high-interaction honeypot in high value targets (e.g. [15, 17, 18, 19]). However, the approach is still not widely adapted in the industry.

Attacker always prefers to hack into the system via Web because many online systems have web application as landing page. There are research works about Web application honeypot (e.g. [16]) to understand the attacker's behavior and try to expose their identities. Other related work includes SDN-based honeypot (e.g. [14]), using stochastic theory to optimize honeypot strategies (e.g. [23]), using game theory to defend the system against strategic attackers (e.g. [24]), and a few others [21, 22]. Most of these approaches have not been verified and tested in real case scenarios. It is not clear if they are effective in practice.

On the other hand, traditional approaches of designing and deploying honeypots may not be very effective now as skilful attackers can identify whether it is a honeypot or not rather easily. To give a simple example, for low-interactive honeypot, if we emulate the system services, the attacker can just enumerate the services by issuing requests. Based on a few pre-fixed responses, an attack can easily confirm that it is a honeypot (See Appendix I for more details). Thus, our approach of designing realistic honeypots in ADIRDS can provide some insights to the community how to design and deploy more realistic honeypots to trap attackers which has been shown to be effective in real case. The work, which is more closely related to our approach is [25], which also proposes to deploy adaptive honeypot. The major difference between our work and their work is that they rely on network status and services to make changes while we are incident response driven.

To summarize, the followings are our contributions:

- We present an incident response framework for zero-downtime online systems, for which it is not possible and impractical to carry out thorough offline forensics study and investigation.
- We illustrate the effectiveness of our proposed solution based on a real case.
- In our proposed solution, we provide a new approach to design realistic honeypots and based on our experiments, we verify that our approach can trap attackers to leave more footprints for further investigation. In our experiment, we are able to capture 38 unique IP addresses from the attackers and identified the attack origin in a shortened period of time during incident response

# 2 Our Approach

## 2.1 Our Attack-Driven IR and Defense Model

### Overview

Figure 1 shows an overview our proposed Attack-Driven Incident Response and Defense System (ADIRDS). There are three major modules: SIEM, Honeypot Deployment Center, and Defense Command Center, together with a novel algorithm called ADIRDM to identify compromised hosts, detect, and analyze threats based on the MITRE ATT&CK matrix, which contains various stages of technical attack tactics and techniques [3]. Our system can support many different platforms such as Windows, macOS, Linux, PRE, Azure AD, Office 365, Google Workspace, SaaS, IaaS, Network, and Containers. Instead of relying on the logs available in the original system, ADIRDS will actively collect evidence from the attack technique perspective while allowing the system to continue to run during the attacks and incidence response. The SIEM will conduct real-time log analysis. Based on the input from incident response, investigation and attack analysis, and correlation in SIEM, we can carry out a Honeypot deployment planning to deploy realistic honeypots to “trap” and “lure” the attacker to leave more footprints to us for further investigation. In addition, Defense Command Center can publish instruction to implement controls and blocking to affected server.

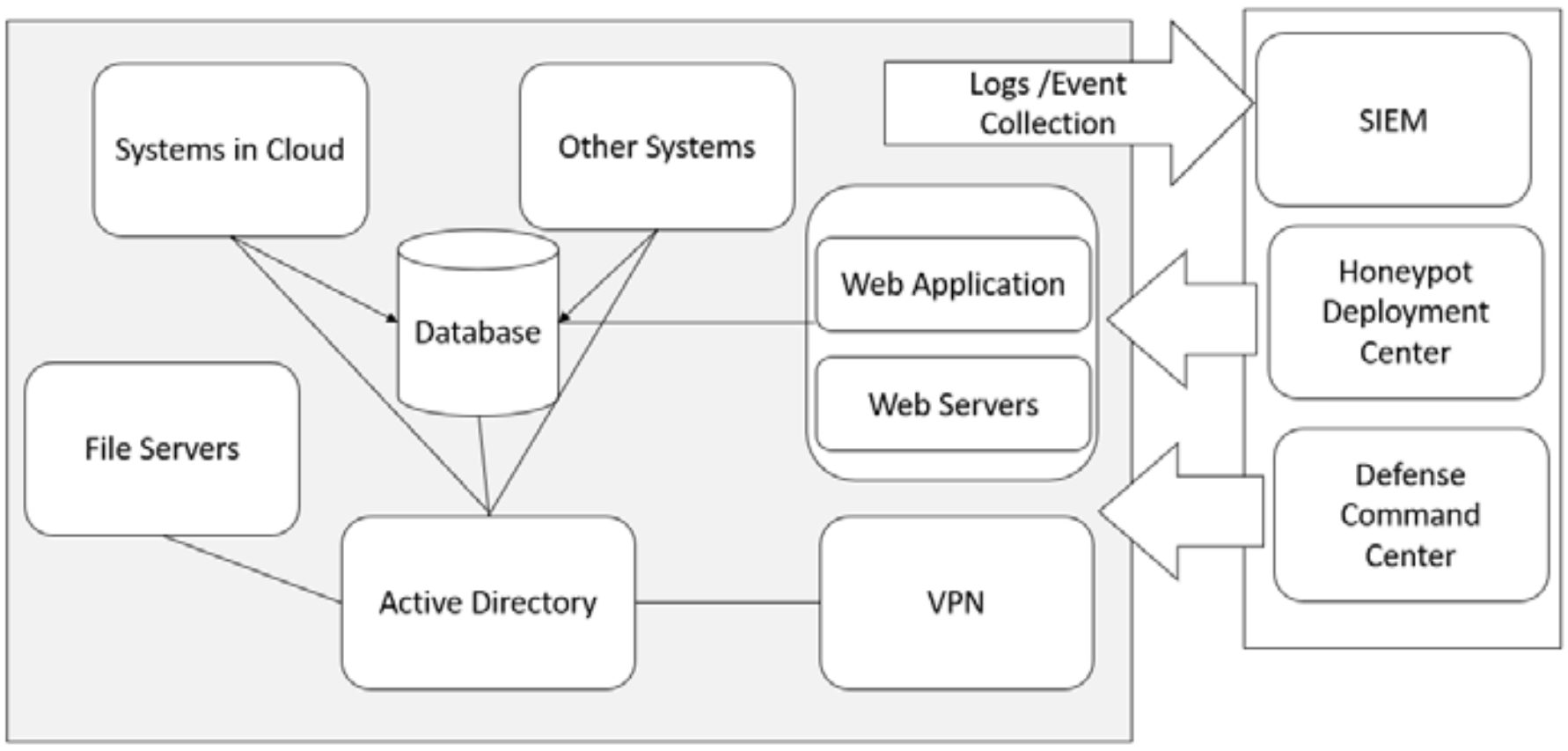

**Figure 1.** An Overview of our Attack-Driven Incident Response and Defense System (ADIRDS)

We select honeypot based on the attacker’s level (see Table 1) and footprints left as well as their interest, thereby attempting to capture more threat intelligence, footprints, and activities of advanced-level attackers as a last resort.

**Table 1.** Levels of Attacker (as known as Blackhats)

| **Level of Attackers** | **Indicators** | | | |
|---|---|---|---|---|
| | **Attacking’s Host IP address** | **Logs / Activities / Configuration/ Scripts footprints** | **Known Vulnerability / Misconfiguration** | **Known Exploit** |
| Novice | Yes | Yes | Yes | Yes |
| Intermediate | Sometimes | Sometimes | Yes | Yes |
| Advanced | No | Mostly No.<br><br>Even yes, the logs or activities are removed after the action is done. | No | No |

## 2.2 Algorithms

We define the network we need to investigate as a graph $G$ as $G = (V, E)$ where $V$ denotes a set of servers, computers, or accounts ($v_1$,$v_2$,$v_3$,…,$v_N$), named as *nodes*. $E$ is a relationship set of connections ($e_1$,$e_2$,$e_3$,…,$e_N$), between pairs of nodes, named as *edges*. We label the blue line as the connection between two trusted systems/parties deemed by the organization. We label the connection as a red line when there is a connection between a trusted system and another untrusted system or between two untrusted systems. We have defined the number of authenticated systems connected to the node $v_i$ as Ns($v_i$) and number of different user accounts used to authenticate to different systems as Nua(Ns($v_i$)). To differentiate a configuration $c_i$ with a legitimate one in a node, we define it as $Di[c_1$,$c_2$,…,$c_N]$.

Attackers attempt to take over each connection or trusted parties to become their stepping-stone to reach out to the final target, in our real case, is the database (see Section 3), blue and green nodes are trusted from the organization's perspective. With this algorithm, we have detected and identified the unknown attack origin effectively.

In Algorithm 1 (see below), we mainly carry out incident response and evidence collection, network connection, accounts, and configuration review. If any of the conditions from a) to d) exists, it will execute Algorithms 2a and 2b. From Algorithm 2a, we will discover and obtain any evidence matching with attack techniques in the ATT&CK matrix and attempt to gather more footprints from the node if possible. From Algorithm 2b, our objectives are to identify defense and mitigation techniques matching with the attack techniques in the matrix for incident containment purposes and discover and collect unmonitored events which are related to the incident. Finally, from Algorithm 2c, we deploy high-interactive Honeypot with consideration of the status of that type of honeypot whether is deployed and compromised, placing accessing information to them to lure the attackers and capture any information whether the Honeypot is compromised. Other than that, practically, we will deploy corresponding firewall rules to limit the inbound and outbound to our honeypots in honeynet only to protect the production from further compromise.

For a more advanced deception strategy to deal with the expert-level attacker, we will implement and monitor [26] different vulnerable Honeypots in docker with the latest and relevant exploits [27] matching with the preferred software, system, and business setting in the affected company and industry. In addition, we can consider generating honeypot and/or honeynet scenarios following the course syllabus of highly technical and industry-recognized hacking professionals [28].

**Algorithm 1.** Attack-Driven Incident Response and Defense Model (ADIRDM)

For every node ($v_i$): start with the least number of trusted connections to the target node:

a. Examine user authentication and Delete-Create-Execute-Delete-Create operations or activities over the files/stored procedures/scripts/user account in the event or/and activity logs in different systems or/and application of the potentially compromised host.
b. Examine the number of authenticated systems connected to the examined node (Ns($v_i$)) AND examine the number of different user accounts used to authenticate to different systems (Nua(Ns($v_i$))).

If the ratio Nua(Ns($v_i$))/Ns($v_i$) > 1, it is suspicious where a single node ($v_i$) is authenticated to many different systems with multiple different user accounts that are not in normal business practice.

c. Differ configuration files of any service available to untrusted parties with the intended and legitimate configuration Di[$c_1$,$c_2$,…,$c_N$].

d. Label the node($v_i$) and edge ($e_i$) connected to and from any node ($v_N$) as red if any item from 1 to 3 is positive and suspicious.

If any red node ($v_i$) satisfies any of the above checkpoints from a) to d):
algorithm_2a($v_i$)
algorithm_2b($v_i$)
If results from algorithm_2a($v_i$) and algorithm_2b($v_i$) are empty:
algorithm_2c($v_i$)

**Algorithm 2a.** Attack-Driven Incident Response and Defense Model (ADIRDM) – Incident Response

Define A is the set of Attack Techniques (A) = {A0, A1, A2, …., An}, where n is a positive integer number from 0 to n, and denote A$v_i$ is the attack technique set of node ($v_i$).

Define E is the set of evidence (E) = {E0, E1, E2, …., En}, where n is a positive integer number from 0 to n, and denote E$v_i$ is the evident set of node($v_i$).

Run check in Atta&k Matrix for node($v_i$):
E$v_i n$ = Obtain corresponding systems/applications process history, logs and activities.

E$v_i(n+1)$ = Differencing the process list, configurations and folder/file list between freshly built server and victim server.

E$v_i(n+2)$= Discover additional unrevealed footprints and evidence from (E$v_i n$ + E$v_i(n+1)$)

**Algorithm 2b.** Attack-Driven Incident Response and Defense Model (ADIRDM) – Defense

Define A is the set of Attack Techniques (A) = {A0, A1, A2, …., An}, where n is a positive integer number from 0 to n, and denote A$v_i$ is the attack technique set of node($v_i$).

Define D is the set of Defense/Mitigation Techniques (D) = {D0, D1, D2, …., Dn} corresponding to the Attack Techniques (A), where n is a positive integer number from 0 to n, and denote D$v_i$ is the attack technique set of node($v_i$).

Run check in Atta&k Matrix for node($v_i$):
A $v_i(n)$ = Identify all services accessible, vulnerabilities and network connections of $node$ ($v_i$) from immediate untrusted and trusted nodes.

Match and Correlate A$v_i(n)$ to D$v_i(n)$ with the highest probability (P) to defense/mitigation in the ATT&CK matrix.

Enable and collect unmonitored service logs at node($v_i$) to SIEM.

**Algorithm 2c.** Attack-Driven Incident Response and Defense Model (ADIRDM) – Honeypot Deployment

H <- Idling server in V or newly added instance to G

Denote $S_i$ as the subset of honeypot being deployed to G in the ith stage
Initialization:
We pick $S_o$ and $S_1$based on difficulties of compromising
And we place the accessing information for $S_1$ inside $S_o$, to encourage attackers to perform lateral movement within our network
While any login attempts appear on the machine within $S_i$ and IR not yet finished:
$C_i$ <- Dictionary for recording the number of honeypots compromised based on nature of the server (e.g., Database, webserver, etc.)
For each compromised honey pot $h_i$:
$C_i$[ $h_i$.type] +=1
$W_i$ <- Array of reciprocal of the ratio of each type of instance in ith stage
$S_{1+2}$ <- A set of V∈H, where not exist in $S_k$ where i+2>k and select randomly based on $W_i$ (i.e. the higher the reciprocal, the higher the chance it appears) + randomly select from the type of machine not appears inside $W_i$
Place the accessing information for $S_{1+2}$ to $S_{1+1}$

To decide how the honeypots should be deployed, we should consider the cost and the benefit of the honeypots. Deploying a honeypot will incur deployment cost and cost of revealing sensitive information to the attackers. However, the presence of honeypots allows the incident response team to have more information and more time to analyze the root cause of the incident. From the game theory perspective, the strategy of the incident responder is to select the level of deployment, and the strategy of the attacker is to decide whether to perform lateral movement. The incident responder should balance between the benefit and the cost when deploying the honeypot.

## 3 Case Study and Walkthrough

Figure 2 depicts a high-level infrastructure diagram of the online gaming company for which our team was appointed to carry out the incident response.

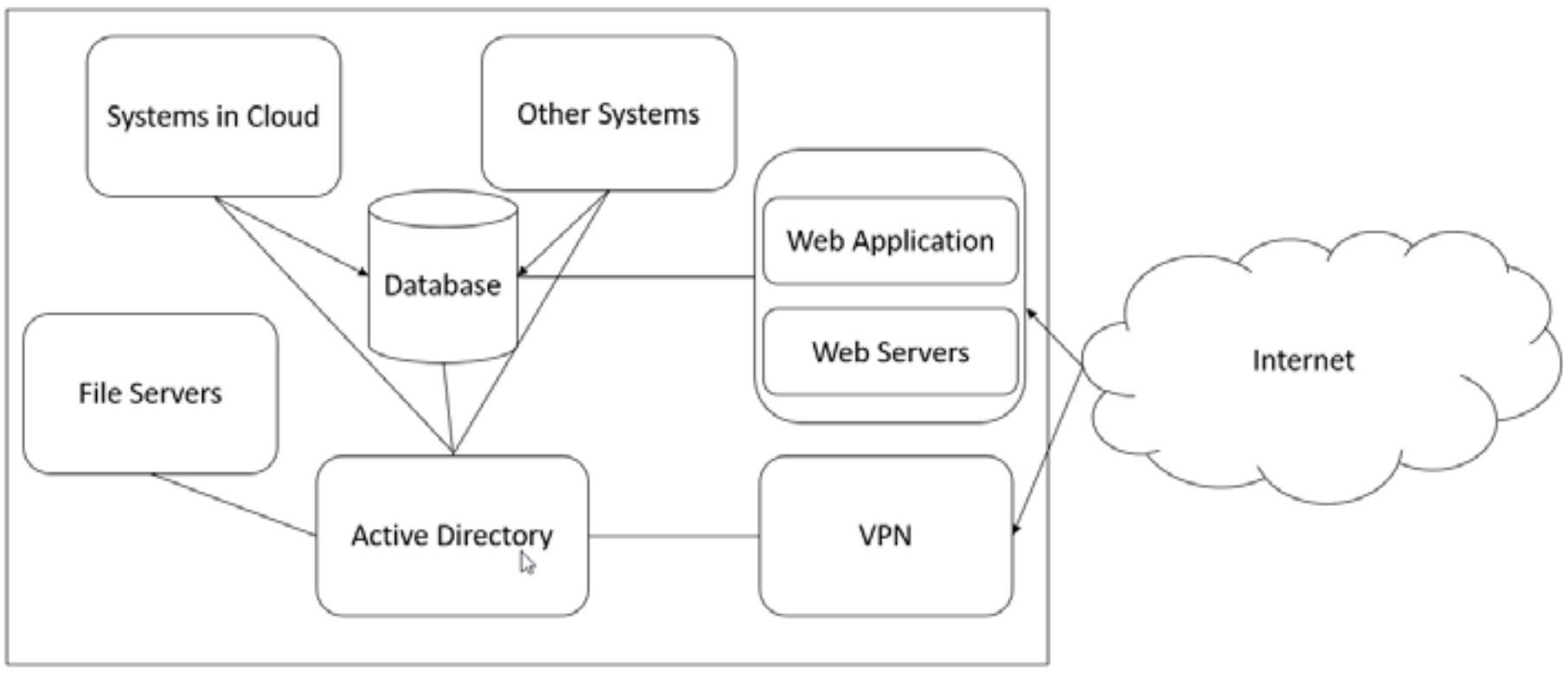

**Figure 2.** High-Level Graph of Node Deployment in Online Gaming Company.

We first carry out a standard IR procedure according to industrial standard. Table 2 shows that two attacks (compromise database transaction records and backdoor discovered from web server) were not successfully resolved using these typical IR procedures. Since the system cannot be shut down, we tried to use our ADIRDS to collect more evidence for these two attacks. In addition, we also cannot identify the attacker's host. And unfortunately, the database is still accessible by the attacker to change the transaction records.

After we establish and apply ADIRDS with the ADIRDM algorithms (1, 2a, and 2b) to each node in the graph (Figure. 2) by adding relevant defensive and mitigation controls including SIEM, Two-Factor Authentication, logs collection, patching server, the shutdown of unnecessary service, etc. We carry out a detailed investigation and comparison between freshly built servers and connected hosts until we have found the Microsoft IIS Web server native module startup failure logs and identified the differences in the configurations. The backdoor is finally detected. Note that all installed anti-virus software and Microsoft Windows Defender cannot detect them this backdoor. We then remove the backdoor, and the database transaction is no longer modified, and the server is still under being monitored in the online gaming company using ADIRDS even after the attacks have been resolved. This real case demonstrates the effectiveness of ADIRDS and show that it can be a solution to incidence response when there is insufficient log information to resolve the attack while the system cannot be shut down to carry out an in-depth incidence response analysis.

**Table 2. Application of Incident Response Strategy**

| **Incident** | **IR methodology** |
|---|---|
| Compromise database transaction records<br><br>It is found the transaction is modified by an attacker by enabling another database to receive the transactions. All database logs and related incoming and outgoing firewall logs are examined but cannot decide the attack root cause. The attacker still can change the transaction during our incident response. | Typical IR does not work. |
| Compromise VPN and Firewall Rules<br><br>The VPN gateway is not updated, and the attacker can dump the credentials of the VPN gateway and access the internal network. 2 Factor Authentication is not implemented. | Typical IR works |
| Backdoor discovered from Web Server<br><br>There is a native module deployed in the Web server which allows a remote attacker to access and dump the credentials and data in the Web server. No logs are available to detect the backdoor installation. | Typical IR does not work |
| Malware spread out in Skype | Typical IR works |
| The skype software is used by the customer service officer. However, attacker fakes the company's skype account to spam their customers. | |

## 4 Experiment on our honeypot deployment strategy

To further demonstrate how to deploy an attack & IR driven, and business context realistic honeypot, we have taken an open-source Web casino system (Figure 3) called Web Poker, then convert it to a honeypot according to our design as a cloud service. This honeypot must be made to be relevant to the target company's business context, according to the design principles we present in the next section to lure the attacker to leave any footprints to our Honeypot system when they are carrying out any reconnaissance and scanning for lateral movement, such that we can discover more artifacts of the attack and incident happened. The duration lasts for 35 days, and we relax the security controls incrementally.

We have included the following rules to promote and deploy our context-related honeypot if authorized by the system owner:

Promotion

- Sharing the IP address(es) of the casino honeypot in some online gaming forums.

Deployment

- Sharing the IP address(es) of the casino honeypot in some network and system configuration files including DNS file and robots.txt.
- Sharing the IP address(es) of the casino honeypot in configuration files in the victim server.
- Setting up a few guessable passwords of administrative and player accounts.
- Open uncommon ports to external and access the system files via directory listing.
- Setting up login sessions of a room such that attackers can see there are players online in a room if they were successfully login.

Log/Event Capture

- Switch on Sysmon and event logs monitoring.
- Capture all fields of HTTP request logs.

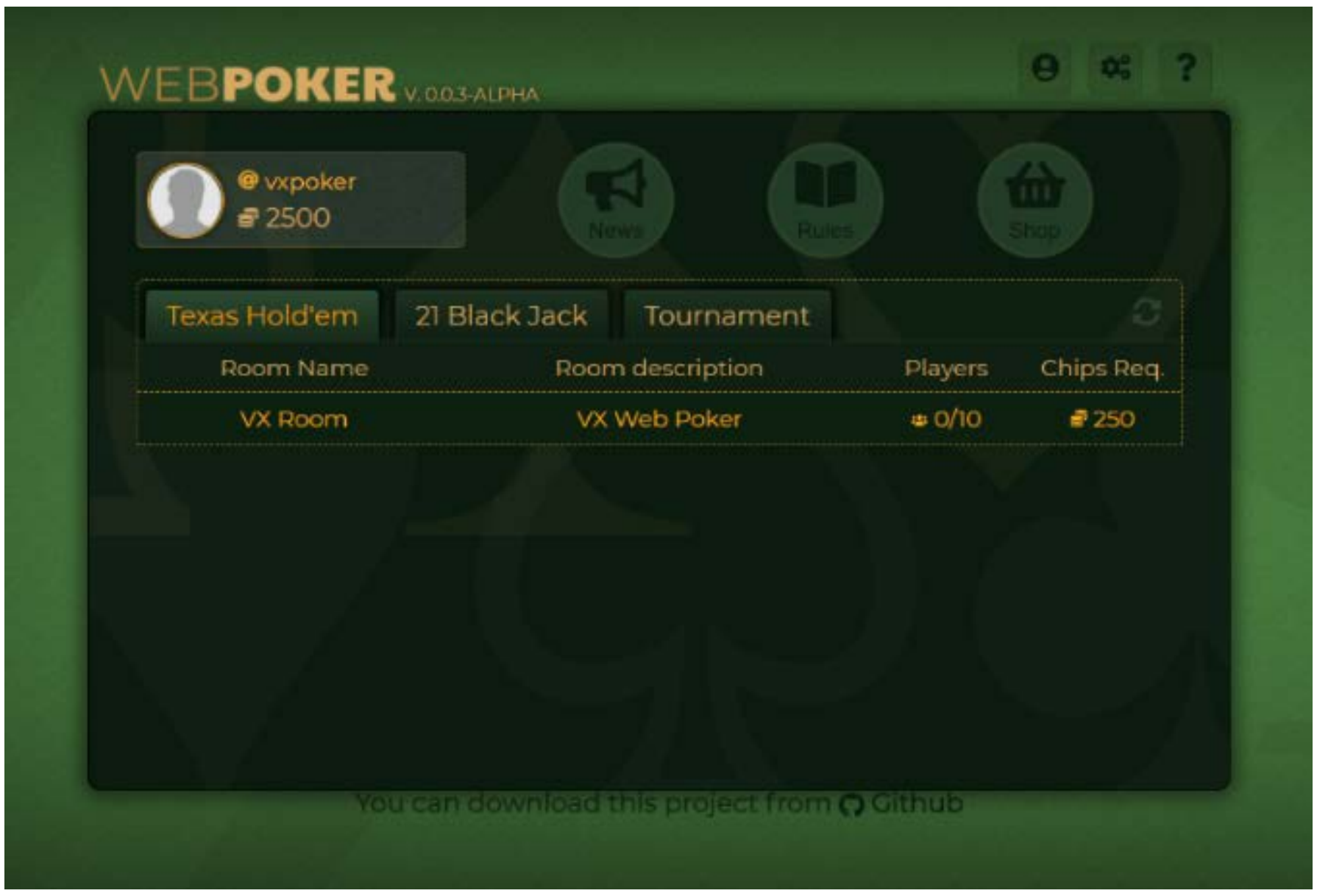

**Figure 3.** High Interaction Business Context-related Webapp Honeypot

We have deployed the honeypot for 35 days with 27,175 access to our honeypot, and there are 38 unique attacker's IP addresses are identified. We have the following with the following incremental deployment plan of our artificial vulnerabilities:

**Table 4a.** Honeypot Deployment Plan

| **Duration** | **Web portal** | **Firewall (allow inbound and outbound traffic)** |
|---|---|---|
| $1^{st}$ - $10^{th}$ day | Deploy typical user accounts with weak passwords<br><br>Keep several rogue players online and login | 80,443 |
| $10^{th}$ – 20th | Deploy admin account with a weak password | 80,443 |
| $21^{st}$ - $35^{th}$ day | Allow directory browsing of the system configuration files under port 9000 | 80,443, 9000 |

Afterward, we examine the logs on daily basis, we have highlighted the top 10 interesting attackers' IP addresses and Correlated attacks and payloads which are shown in the table:

**Table 4b.** Attack logs analysis of High-interaction Business Context-related Honeypot

| Duration | IP address | Attack No. of Attempt & Duration | Payload/Activities |
|---|---|---|---|
| 1st - 10th day | 167.99.8.241 | 64, 1 min | Looking for sharing folder, configuration files, DB files, and upload folder |
| | 5.188.210.227 | 21, 1 min | Looking for PHP admin page and echo.php |
| | 45.146.164.110 | 500, 5 mins | Looking for, ThinkPHP, WordPress login and admin page |
| | 84.152.64.124:4444 | 6, 30s | CONNECT request to the web application |
| | 223.247.179.82 | 3, 10s | Get webapp user configuration files |
| | 66.240.192.138 | 6, 4 mins | 1. First Visit<br>2 .Read robots.txt<br>3. Looking for sitemap.xml<br>4. Try to GET /.well-known/security.txt but failed to open /usr/share/nginx/html/.well-known/security.txt<br>5. Launching unknown exploit code<br>Example:<br>"\x03\x1F@Ba\x00\x00\x00\x00\x00\x00\x00\x00\x00\x00\x00\x00\x00\x00\x00\x00\x00\x00\x00\x00\x00\x00\x00\x00\x00\x00\x00\x00\x00\x00\x00\x00\x00\x00\x00\x00\x00\x00\x00\x00\x00\x00\x00 [..] |
| 10th – 20th | 35.179.93.71 | 2, 2 mins | Scanning with open source scanning tool:<br>Example:<br>"GET / HTTP/1.0" 200 1219 "-" "masscan/1.3 (https://github.com/robertdavidgraham/masscan)" "-" |
| | 161.35.104.71<br>128.199.22.35 | 182, 5 mins | Launching PHP Exploit and looking for PHP configuration files<br>Example:<br>"GET /index.php/PHP%0Ais_the_shittiest_lang.php?QQQQQQQQQQQQQQQQQQQQQQQ [….] |
| | 58.65.163.89 | 40, 30s | Scan SQLite admin |
| 21st - 35th day | 185.100.86.128 | 1, 3 mins | Download the configuration, docker file, source code in source directory. Example logs:<br>`185.100.86.128 - - [19/Oct/2021 04:25:46]` |
| | | | "GET /web-poker/.git/info/exclude HTTP/1.1" 200 -<br>185.100.86.128 - - [19/Oct/2021 04:26:01] "GET /docker-compose.yml HTTP/1.1" 200 -<br>185.100.86.128 - - [19/Oct/2021 04:26:17] "GET /web-poker/ HTTP/1.1" 200 -<br>185.100.86.128 - - [19/Oct/2021 04:26:20] "GET /web-poker/frontend/ HTTP/1.1" 200 -<br>185.100.86.128 - - [19/Oct/2021 04:26:31] "GET /web-poker/backend/ HTTP/1.1" 200 -<br>185.100.86.128 - - [19/Oct/2021 04:26:36] "GET /web-poker/backend/orchestrator/ HTTP/1.1" 200 -<br>185.100.86.128 - - [19/Oct/2021 04:26:40] "GET /web-poker/backend/orchestrator/Dockerfile HTTP/1.1" 200 - |

We take those Attackers' IP addresses, correlated attack activities, and/or payload as the parameter to query all system logs and detect any of them are manipulated by the attacker.

As a threat intelligence, those IP addresses can be taken as pre-alert to the system administrator such that he can conduct preventive countermeasures to block those malicious scanner IP addresses at Firewall Level. We are particularly interested in the IP addresses launching the exploit and downloading the code and configurations. We are taking and passing them to the following artifact retrieval algorithm (Algorithm 3) and making a further defense for the systems. For example, we have found that the 185.100.86.128 is from a TOR entry and exit, thereby we can consider blocking the traffic from the TOR network and blocking the IP address (66.240.192.138) who is a more determined and advanced attacker who launched an unknown exploit.

**Algorithm 3.** Artifact Retrieval and Incident Containment Algorithm

```
For each IP address (ip) captured in Honeypot:
        Query Server System/Event Logs (ip)
        Query storage event logs(ip)
        Query VPN logon/logff event(ip)
        Query Email logon/logoff event(ip)
        Query Firewall Configuration(ip)
        Query Firewall Rules (ip)
```

| |
|---|
| Query VirusTotal (ip)<br>Scanning TOR Entry or/and Exit (ip)<br>If any of them return true with malicious indicators:<br>- Block Firewall Rules (ip)<br>- Contain and/or isolate any machine with outbound traffic (ip) |

We have dealt with another incident in the same company**.** As we have deployed low interactive, high interactive and realistic honeypot for comparison and evaluation whether we can identify the attacker's attack vector and origin (Table 4c). We can successfully lure the attacker to download our mirrored configuration files for their lateral movement. Meanwhile, we can identify their origin from another compromised contracted vendor workstation via remote desktop connection with weak password. Low and high interaction honeypots can capture scanning traffic; however, we cannot clearly find out the attacker's origin while our realistic honey can achieve what we want to do within 48 hours

**Table 4c.** Evaluation between attacks capture among Low Interaction, High Interaction and Realistic Honeypots

| Attack Activities | Low Interactive Honeypot | Frequency (Within 48 hours) | High Interactive Honeypot | Frequency (Within 48 hours) | Realistic Honeypot | Frequency (Within 48 hours) |
|---|---|---|---|---|---|---|
| **Capture Exploit and Vulnerability Scan Traffic** | yes | 7893 | yes | 7812 | yes | 5478 |
| **Capture Web Attack Payload** | Yes | 767 | Yes | 784 | Yes | 779 |
| **Capture SSH attack** | Yes | 690 | Yes | 662 | Yes | 672 |
| **Attacker Revisit** | yes | 8232 | No | 7123 | Yes | 1704 |
| **Attacker Download our deceptive but mirrored configuration files via directory traversal** | No | 0 | No | 0 | Yes | 13 files downloaded. Attacker IP captured |
| **Attacker Attempts to Login to Our Web application honeypot** | No | 0 | No | 0 | Yes | 125 |
| **Attackers attempt to brute force password attack of ourweb application honeypot** | No | 0 | yes, it is not related to the context of the web application | 14 | Yes | 6 times success / 125 times |

# 5 Design and Limitations of Realistic Honeypot

## 5.1 The Design

In this section, we briefly talk about how to design a more realistic honeypot to trap attackers. The followings show some of the key design issues. We have attached an Appendix I for comparison between typical honeypot and our proposed realistic honeypot.

(1) We use real services and ports (but use emulation to do it).
(2) We need to clone the real or similar system to be the honeypot.
(3) Fake data needs to be generated to make it look real.
(4) We deploy the honeypot in real physical machine to avoid being discovered by the attacker
(5) We need to deploy to the same network subnet to make it look real, e.g., as a DEV or UAT system.
(6) We simulate the number of users (with different IP addresses) logging into the system so that the attacker can see how many users currently logon to the system.
(7) We deploy reasonable vulnerabilities according to OWASP Top 10 and SANS top 25 vulnerabilities.
(8) We use similar naming convention and same account names as in the real system in the honeypot.
(9) We show them that we have done some hardening like server banner removal, patches on several libraries to make it look like a real system.
(10) We make sure that logs can be exported via different means including SSH and tunneling over other protocols to get our logs.

To protect the real system, we need to do the followings:
(1) Never connect to the production server.
(2) We need to have reborn the machine and export the logs in stealthy way.
(3) It is better to deploy the kernel driver to capture the attacker’s activities if possible.

## 5.2 Challenges and limitations

To produce such a realistic honeypot, we need to face the following challenges.
(1) We need to manually review the existing network and system of the target company.
(2) We need to select related applications and network devices, servers like the target system.
(3) We need to set up and configure the system similar to the real system.
(4) We need to deploy the honeypot on the same network or with the same network service provider of the target system.
(5) We need to enable shared drive/service but need to link them to another honeypot like database
(6) We need to install monitoring service on top of the system

Thus, a future direction is how to automate this process as much as we can.

## 6 Future Work and Conclusions

For future research, we can consider the automation of Honeypot deployment on the fly with consideration of the configurations of the victim servers, attacker's footprint, logs creation, and file/folder change, customizing a more realistic dynamic environment such that the attacker is confused to expose their footprints instead of manual deployment.

We should facilitate and customize the high interaction honeypot with authenticity, flexible deployment, ease of operation, support scalability, realistic high-interaction, attack pivot point, and deception credentials. Making the honeypot become more realistic and confusing the attacker is very challenging, we will deploy different combinations of honeypot in VM, docker, real physical machine, and a cloud-based server.

Concerning the collected various attack logs, other researchers discuss over-optimizing honeypot deployment strategy with various algorithms with limited attacker information [23], we can further customize our Honeypot strategies in different periods (p) to set up their "preferred" vulnerable environment and server for their further intrusion. The restoration and logs monitoring of the honeypot is essential to collect as many logs as possible and maintain the uptime of the honeypot. In the honeypot logs capturing perspective, we consider a more stealthy and low-level approach to capture the logs and network traffic. Taking the latest vulnerable scenario and exploit proof of concepts are important to create a realistic honeypot, maximizing the probability of the attacker to approach the honeypot. In addition, we plan to carry out an extensive experiment for evaluation.

## Appendix I : Typical Honeypot Vs Realistic Honeypot

| **Nature** | **Typical Honeypot** | **How attacker detect honeypot?** | **Realistic honeypot** |
|---|---|---|---|
| Emulation of service | Emulate system services for low-interaction honeypot. | Attacker can enumerate the service by issuing requests, as honeypot always return a few fixed responses and cannot give additional response if the attacker changed their requests. If the attacker is an automatic bot /scanner, it is okay. But if the attacker is a human operator, it will be revealed. | We use real services and ports but emulation. |
| Real service running | Real service and ports, for high interaction honeypot. | Attacker can find out whether it is under VMware and driver adapter, and the windows size of the packet, need to match the platform profile of other production servers. If other servers are not VMs, the honeypot must not be in VM. if the customer uses Win10, honeypot should not use Win7. Meanwhile, honeypot in general exposes too many services to let the attacker to exploit. Honeypot mainly is for network probing by attackers; however, attacker simply takes one more step to enumerate the service of the honeypot, it is found that it is not intended for business purpose. Meanwhile, it is rare to see application honeypot with look-alike real business users / data. | Yes, we aligned with high interaction honeypot with real services. Meanwhile, we will install the application and tools/software which are mirrored from other production and testing systems of the company. |
| Real system | Not necessarily installed with mirrored systems to avoid the risk of information leakage. | Attacker is easily finding out the system is not related to company business and operation if the configuration and system version is not aligned with other systems in the same network. | We mirror and clone the real or similar system configuration as honeypot to confuse the attacker. |
| Database data | Not necessarily deployed. | Most database honeypot is with fake data. | We generate data by reference to any testing |
| | | | data database records of the company. |
| Simulation of logon users | Not included. | Typical honeypot (low and high interaction honeypots) does not support application-level logon simulation). | We will simulate number of users logging into the system. Attacker can see how many current user logons to the system.<br><br>We will simulate users logging in from different IP addresses. |
| Vulnerability introduction and deployment | Service can be with vulnerability, however most honeypot deploys "too many" vulnerabilities like a playground, hacker will suspect it is a honeypot as it is too easy. | We cannot introduce too old vulnerabilities and must match the technical platforms of the customer target site. The vulnerabilities are not update to date; we need a distance D to calculate. For example, D is the vulnerability distance between the latest update and vulnerable version of system service, we should keep it as short as possible. For example, you cannot introduce a RCE (Remote Code Execution) vulnerabilities which is 10 years ago, however, it does not happy in the same platform of another machine. | We reference and deploy reasonable vulnerabilities according to OWASP Top 10 and SANS Top 25 selectively. |
| User Accounts set up | It may be different from production to avoid information leak and further compromise. | Attacker will feel suspicious about the account naming and ID convention are completely different from their gathered or compromised accounts. | We will use the similar naming convention and same account name for the systems. |
| Hardening | Not necessarily. | Honeypot services can be emulated or real one. However, it is not common to deploy vulnerability patch to make the attacker believe it is a realistic one. | We will show we have done some hardening like server banner removal and keep several libraries are patched. |